%% file: main.tex
\documentclass[sigconf]{acmart}
\AtBeginDocument{%
  }

\usepackage{xcolor}
\usepackage[table,xcdraw]{xcolor}
\usepackage{progressbar}
\usepackage{tikz}
\usepackage{adjustbox}
\usepackage[most]{tcolorbox}
\usepackage{fontawesome}
\usepackage{multirow}
\usepackage{booktabs}
\usepackage{makecell}
\usepackage[normalem]{ulem}
\usepackage[table,xcdraw]{xcolor}
\usepackage{tabularray}
\definecolor{Gallery}{rgb}{0.937,0.937,0.937}
\usepackage{colortbl}
\usepackage{tikz}
\usepackage{array}
\usepackage[table,xcdraw]{xcolor}
\usepackage{soul}
\usepackage[table,xcdraw]{xcolor}
\usepackage[para]{footmisc}
\usepackage{enumitem}
\usepackage{stfloats}
\usepackage{arydshln}
\definecolor{darkgreen}{RGB}{0,120,0}
\usepackage{placeins}

\usepackage[utf8]{inputenc}
\usepackage{geometry} % Only if not already loaded by the class
\usepackage{tabularx}

\usepackage{soul}
\usepackage{xcolor}

\newcolumntype{P}[1]{>{\raggedright\arraybackslash}p{#1}}

\newlist{tabitem}{itemize}{1}
\setlist[tabitem]{label=---, nosep, leftmargin=*, before=\vspace{-0.5em}, after=\vspace{-1em}}

\usepackage{array}

\usepackage{tcolorbox}
\newtcbox{\step}{on line, colframe=black, colback=white, boxrule=0.8pt,
  arc=0pt, boxsep=1pt, left=2pt, right=2pt, top=0pt, bottom=0pt}

\definecolor{darkgreen}{rgb}{0.0, 0.6, 0.2} % Adjusted to be brighter and more distinguishable
\definecolor{lightgreen}{rgb}{0.3, 0.5, 0.4} % Slightly darker and closer to reddish-orange

\newif\ifuseMicrosoft
\useMicrosoftfalse  % Use Company X

\ifuseMicrosoft
  
\else
  
\fi

\usetikzlibrary{fit} % To automatically adjust the size of the border

\copyrightyear{2026}
\acmYear{2026}
\setcopyright{none}
\acmDOI{XXXXXXX.XXXXXXX}
\acmConference[FSE'26]{Companion Proceedings of the 34th ACM Symposium on the Foundations of Software Engineering}{June 5--9, 2026}{Montreal, Canada}
\acmBooktitle{Companion Proceedings of the 34th ACM Symposium on the Foundations of Software Engineering (FSE '26), June 5--9, 2026, Montreal, Canada}
\acmISBN{XXX-XXXX/XXXXXXXX}

\newcommand{\boldification}[1]{\ifdraft\indent ** \textbf{#1}** \\ \indent\else\relax\fi}
\newif\ifdraft
 \draftfalse %hides boldifications
\begin{document}

\newcommand{\votingprogressbar}[3]{
  \begin{tikzpicture}[baseline]
    % Position the Minimum label on the left
    \node[anchor=west, inner sep=0pt] (minlabel) at (0,0) {#1};
    
    % Draw the progress bar with minimal width and spacing
    \node[anchor=west, xshift=0.2em] (bar) at (minlabel.east) {\progressbar[linecolor=red!60, filledcolor=red!60, width=1.5cm]{#2}};
    
    % Position the Maximum label on the right
    \node[anchor=west, xshift=0.2em] (maxlabel) at (bar.east) {#3};
  \end{tikzpicture}
}

\newcommand{\customprogressbar}[3]{
  \begin{tikzpicture}[baseline]
    % Draw the border dynamically around the entire content
    \node[draw=black, inner sep=0.3pt, rounded corners, fit={(0,0) (5,0.5)}, thick] (border) {
      \begin{tikzpicture}
        % Position the Minimum label on the left
        \node[anchor=west] (minlabel) at (0,0) {#1};
        
        % Draw the progress bar in the center
        \node[anchor=west] (bar) at ([xshift=1em]minlabel.east) {\progressbar[linecolor=blue, filledcolor=green]{#2}};
        
        % Position the Maximum label on the right
        \node[anchor=west] (maxlabel) at ([xshift=1em]bar.east) {#3};
      \end{tikzpicture}
    };
  \end{tikzpicture}
}

\newcommand{\AS}[1]{\textcolor{red!60!red}{#1}}

\newcommand{\GW}{\textsc{glue work}}

\newcommand{\percentagebar}[1]{%
    \begin{tikzpicture}
        % Draw the outer rectangle (full width) with dark gray background
        \draw[fill=darkgray] (0,0) rectangle (2,0.3);
        % Draw the inner rectangle (progress part) with dark green fill
        \fill[darkgreen] (0,0) rectangle (#1*2/100,0.3);
        % Add the percentage text in white and bold
        \node[text=white, font=\bfseries] at (1,0.15) {\small #1\%};
    \end{tikzpicture}%
}

\newcommand{\percentagebarrr}[1]{%
    \begin{tikzpicture}
        % Draw the outer rectangle (full width) with dark gray background
        \draw[fill=darkgray] (0,0) rectangle (2,0.3);
        % Draw the inner rectangle (progress part) with dark green fill
        \fill[lightgreen] (0,0) rectangle (#1*2/100,0.3);
        % Add the percentage text in white and bold
        \node[text=white, font=\bfseries] at (1,0.15) {\small #1\%};
    \end{tikzpicture}%
}

\newcommand{\ossnabox}{\fcolorbox{darkgray}{lightgreen}{\textbf{\textcolor{white}{OSSNA}}}}

\newcommand{\fossbox}{\fcolorbox{darkgray}{darkgreen}{\textbf{\textcolor{white}{FOSSAsia}}}}

\definecolor{CategoryColor}{HTML}{4C78A8} % change to match your scheme

\newcommand{\barChart}[1]{%
  \begin{tikzpicture}[baseline=(current bounding box.center)]
    % Scale bar length (narrow: max 1.5 cm for 100%)
    \pgfmathsetmacro{\barlen}{#1*0.015}
    % Bar (slim, no border)
    \draw[fill={rgb,255:red,119;green,207;blue,166}] (0,0) rectangle (\barlen,0.25);

    % Label on the side, aligned to bar end
    \node[font=\normalsize\bfseries,black,anchor=west] at (\barlen+0.1,0.125) {#1\%};
  \end{tikzpicture}%
}

\newtcolorbox{bubble}[1][]{colback=blue!5!white, colframe=blue!50!white, fonttitle=\bfseries, title=#1, width=\linewidth, boxsep=1pt, top=3pt, bottom=4pt, before skip=6pt, after skip=4pt, left=4pt, right=4pt}

% Vertical SPACE bubble: Positive (top) → Neutral/Negative (bottom)
\newtcolorbox{spacebubble}[2][]{%
  enhanced,
  breakable,
  colback=black!2!white,
  colframe=black!40!white,
  boxrule=0.6pt,
  arc=1.5mm,
  width=\linewidth,
  boxsep=1pt,
  top=3pt,
  bottom=3pt,
  left=3pt,
  right=3pt,
  before skip=5pt,
  after skip=5pt,
  fonttitle=\bfseries,
  title={#2},
  #1
}

% Positive section (green)
\definecolor{spaceblue}{HTML}{0D92F4}

\newtcolorbox{spacepositive}{%
  colback=spaceblue!6!white,
  colframe=spaceblue!50!black,
  boxrule=0.4pt,
  arc=1.5mm,
  boxsep=1pt,
  top=3pt,
  bottom=3pt,
  left=4pt,
  right=4pt
}

% \newtcolorbox{spacepositive}{%
%   colback=green!6!white,
%   colframe=green!50!black,
%   boxrule=0.4pt,
%   arc=1.5mm,
%   boxsep=1pt,
%   top=3pt,
%   bottom=3pt,
%   left=4pt,
%   right=4pt
% }

% Neutral / Negative section (red)
\definecolor{spacered}{HTML}{C62E2E}

\newtcolorbox{spacenegative}{%
  colback=spacered!6!white,
  colframe=spacered!50!black,
  boxrule=0.4pt,
  arc=1.5mm,
  boxsep=1pt,
  top=3pt,
  bottom=3pt,
  left=4pt,
  right=4pt
}

% \newtcolorbox{spacenegative}{%
%   colback=red!6!white,
%   colframe=red!55!black,
%   boxrule=0.4pt,
%   arc=1.5mm,
%   boxsep=1pt,
%   top=3pt,
%   bottom=3pt,
%   left=4pt,
%   right=4pt
% }

% Compact bullet helper
\newcommand{\bitem}{\par\noindent\textbullet\ }

\setlength{\textfloatsep}{5pt plus 1.0pt minus 10.0pt}
\setlength{\floatsep}{5pt plus 1.0pt minus 10.0pt}
\setlength{\intextsep}{5pt plus 1.0pt minus 10.0pt}

\let\svthefootnote\thefootnote
\newcommand\blankfootnote[1]{%
  \let\thefootnote\relax\footnotetext{#1}%
  \let\thefootnote\svthefootnote%
}

% Usage examples:
% \singlebar{75}{blue!70}
% \singlebar{45}{red!60}
% \singlebar{90}{} % Uses default blue color

\tcbset{
    mybox/.style={
        colback=gray!10,   % Background color
        colframe=gray!10,  % Match background to hide frame
        width=\linewidth,  
        boxrule=0mm,       % No border
        leftrule=0mm,      
        rightrule=0mm,     
        toprule=0mm,       
        bottomrule=0mm,    
        sharp corners,     
        boxsep=0pt,        
        left=0pt,          
        right=0pt,         
        top=2pt,           
        bottom=2pt         
    }
}

%%
%% The "title" command has an optional parameter,
%% allowing the author to define a "short title" to be used in page headers.
% \title{Software Engineers Are from Mars, Domain Experts Are from Jupiter: Navigating Interdisciplinary Collaboration Frictions in Developing Software}

% Not Just Faster, But Better
\title{\textit{The Fast and Spurious}: Developer Productivity with GenAI} 

\author{Sadia Afroz}
\affiliation{%
  \institution{Oregon State University}
  \country{USA}
}
\email{afrozs@oregonstate.edu}

\author{Zixuan Feng}
\authornote{Co-first author}
\affiliation{%
  \institution{Oregon State University}
  \country{USA}
}
\email{fengzi@oregonstate.edu}

\author{Tyler Menezes}
\affiliation{%
  \institution{CodeDay}
  \country{USA}
}
\email{tylermenezes@codeday.org}

\author{Katie Kimura}
\affiliation{%
  \institution{Oregon State University}
  \country{USA}
}
\email{kimuraka@oregonstate.edu}

\author{Bianca Trinkenreich}
\affiliation{%
  \institution{Colorado State University}
  \country{USA}
}
\email{bianca.trinkenreich@colostate.edu}

\author{Igor Steinmacher}
\affiliation{%
  \institution{Northern Arizona University}
  \country{USA}
}
\email{igor.steinmacher@nau.edu}

\author{Anita Sarma}
\affiliation{%
  \institution{Oregon State University}
  \country{USA}
}
\email{Anita.Sarma@oregonstate.edu}

\begin{abstract}

Generative AI (GenAI) tools are increasingly being adopted in software development as productivity aids, since there is evidence that GenAI tools can improve individual aspects of productivity. However, productivity is multidimensional; accelerating one aspect of work may simply shift effort to another. In this paper, we investigate how GenAI adoption affects different dimensions of developer productivity. We surveyed 415 software practitioners to understand how they perceive productivity changes associated with AI adoption, using the SPACE framework (\textbf{S}atisfaction and well-being, \textbf{P}erformance, \textbf{A}ctivity, \textbf{C}ommunication and collaboration, and \textbf{E}fficiency and flow). Our results reveal systematic redistribution of effort across SPACE dimensions. While frequent GenAI users reported faster task completion and higher output volume, these gains were offset by increased code review burden, persistent cognitive load from output verification, and unchanged collaboration patterns. We further provide an empirical mapping between the challenges perceived by developers and potential strategies to mitigate them. Overall, our findings suggest that, at the current stage of GenAI adoption, perceived productivity gains may be \emph{spurious}—surface-level acceleration, often accompanied by redistributed effort and hidden costs.

\end{abstract}

%aware of the less glamorous - and often less-promotable -
%Today’s Open Source Software (OSS) communities depend on diverse forms of contributions for long-term sustainability.
%
%However, many OSS projects prioritize core code contributions, often overlooking contributions such as project coordination, documentation, leadership, and mentoring—collectively serving as the glue that builds communities together.
%
%This so-called \GW{}, which is critical to project health, yet remains underrecognized and unacknowledged.
%

%%
%% The code below is generated by the tool at http://dl.acm.org/ccs.cfm.
%% Please copy and paste the code instead of the example below.
%%
%\begin{CCSXML}
%<ccs2012>
%   <concept>
%       <concept_id>10003120.10003130.10011762</concept_id>
%       <concept_desc>Human-centered computing~Empirical studies in collaborative and social computing</concept_desc>
%       <concept_significance>300</concept_significance>
%       </concept>
% </ccs2012>
%\end{CCSXML}
%
%\ccsdesc[300]{Human-centered computing~Empirical studies in collaborative and social computing}

%\ccsdesc[100]{Human-centered computing}
%%
%% Keywords. The author(s) should pick words that accurately describe
%% the work being presented. Separate the keywords with commas.
%% A "teaser" image appears between the author and affiliation
%% information and the body of the document, and typically spans the
%% page.

%\received{20 February 2007}
%\received[revised]{12 March 2009}
%\received[accepted]{5 June 2009}

\maketitle

\input{sections/1_intro}
\label{sec:intro}

\input{sections/2_background_SPACE}

\input{Tables/survey_mapping}
\input{sections/3_Methodology}
\label{sec:method}

\input{sections/4_Results}
\vspace{-5pt}
\input{sections/5_Threats_to_validity}
\label{sec:threats}

\input{sections/6_discussion}
\label{sec:discussion}

 \vspace{-5pt}
\bibliographystyle{ACM-Reference-Format}
\bibliography{bib}
\end{document}

%% file: sections/1_intro.tex
\vspace{-2mm}
\section{Introduction}
Generative AI (GenAI) tools have experienced rapid growth, with software development emerging as an early adopter~\cite{zakharov2025ai,nguyen2025generative}. According to the 2025 Stack Overflow Developer Survey, nearly 84\% of developers use or plan to use GenAI tools \cite{Shimel2025AIAdoption}. This rapid adoption is driven by a widespread belief that GenAI can boost developer productivity \cite{yu2025paradigm}, yet there is limited evidence to understand if GenAI improves the overall productivity, or merely creates a spurious sense of speed.
%the question: \textit{Does GenAI improve overall developer productivity, or merely create a spurious sense of speed?}

To ensure that any observed productivity gains are real and sustainable, it is necessary to consider how GenAI integrates with software engineering (SE) practices. % such as requirements engineering, testing, and governance. 
Without clear process safeguards, GenAI may shift work effort without reducing it---or worse, introduce new forms of technical debt and developer burnout. For example, \citet{moreschini2026evolution} showed that GenAI can incur prompt engineering debt and explainability debt, leaving teams with code that may ``work'' but lacks clarity, testability, or adaptability. Similarly, \citet{feng2025gains} demonstrated that GenAI adoption can heighten burnout by increasing job demands on developers. Thus, achieving productivity gains with GenAI requires alignment with established SE rigor and empirical investigation. In this context, integrating GenAI into software development is not just a tooling decision, but a socio-technical design challenge~\cite{feng2025charting}.

The evidence of GenAI's impact on productivity is mixed. While GenAI tools such as GitHub Copilot and ChatGPT have been reported to automate routine tasks and improve efficiency \cite{peng2023impact, ziegler2022productivity}, \citet{tong2025slow} found that developers completed tasks 19\% slower when using AI, and \citet{vaithilingam2022expectation} observed higher task-failure rates and no significant improvement in completion time. Such inconsistencies likely arise because existing studies often scope their study to particular development activities or narrow productivity metrics rather than taking a comprehensive view.

However, productivity in software development is a multi-dimensional socio-technical system \cite{forsgren2021space}. It is not simply measurable by the output volume. It includes long-term work quality, psychological well-being, team collaboration, uninterrupted focus, and satisfaction with work \cite{storey2019towards, forsgren2021space, rodriguez2012empirical}. When organizations optimize one dimension, they may inadvertently create other constraints. Understanding these inter-dependencies is critical for sustainable GenAI adoption.

To investigate this socio-technical interplay, we adopt the multidimensional productivity framework SPACE \cite{forsgren2021space} as our analytical lens for investigating developer productivity across five dimensions: \textit{Satisfaction and well-being}, \textit{Performance}, \textit{Activity}, \textit{Communication and collaboration}, and \textit{Efficiency and flow}. We employ SPACE as a measurement taxonomy and a structuring lens to understand how perceived productivity gains in one dimension may introduce new demands or trade-offs in others. This systems-oriented perspective helps characterize how developers’ effort and attention shift during GenAI adoption, pinpoint where challenges emerge, and inform strategies for designing more resilient software engineering practices for AI-infused workflows. This motivates us to answer the following research questions:

\begin{itemize}
    \item \textit{RQ1. How does GenAI adoption affect developer productivity across multiple dimensions?}
    \item \textit{RQ2. What productivity-related gaps, challenges, and strategies do developers perceive in GenAI adoption?}
\end{itemize}

% \textbf{\textit{RQ1. How does GenAI adoption affect developer productivity across multiple dimensions?}}

% \textbf{\textit{RQ2. What productivity-related gaps, challenges, and strategies do developers perceive in GenAI adoption?}}

To address these research questions, we conducted a large-scale survey of \textbf{415 professional developers} grounded in the SPACE framework. We analyzed the data using a mix of quantitative and qualitative methods.

As a result, our study situates developers' experience with GenAI within the \emph{SPACE} framework to provide emerging empirical results that raise important discussion questions about what it means to be ``fast'' in the context of broader software development, and the potential for perceived "spurious" forms of productivity in AI-mediated software development. \textit{``I think in the current state, AI can give you a productivity boost by bringing you 70\% there in a few minutes. However, the last 30\% will likely take you close to what you would've done before because you also have to thoroughly review AI code.''} [P20].

Our contributions are twofold. First, combining quantitative patterns with qualitative explanations, we characterize how developers perceive GenAI's productivity impact across five dimensions. Overall, developers do not report substantial productivity changes, although frequent GenAI users perceive modest improvements in \textit{Efficiency and flow} and \textit{Satisfaction and well-being}, with limited perceived gains in \textit{Performance}, \textit{Activity}, or \textit{Communication and collaboration}. Second, we develop an empirical mapping between the productivity-related challenges developers encounter and the strategies proposed to address them, providing actionable guidance for practitioners integrating GenAI into their workflows.

% Our study provides insights for where effort moves within SPACE under GenAI adoption, enabling software engineering teams and GenAI tool developers to more effectively predict and control for shifting effort. 

% According to the findings from quantitative analysis, developers do not perceive that GenAI meaningfully changes productivity across the five \emph{SPACE} dimensions. Compared to developers who use GenAI less frequently, frequent genAI users reported slightly higher perceived improvements in \textit{Efficiency and Flow} and \textit{Satisfaction and Well-being}, but these gains did not extend to \textit{Performance}, \textit{Activity}, or \textit{Communication and Collaboration}. Subsequently, qualitative results provided us an empirical mapping between challenges and practitioner strategies to address the challenges. 
% Our study situates developers' experiences with GenAI within the \emph{SPACE} framework to provide emerging empirical results that raise important discussion questions about being ``fast'' means in the context of broader software development and the potential for perceived "spurious" forms of productivity in AI-mediated software development. 

%% file: sections/2_background_SPACE.tex
\vspace{-3mm}
\section{Background and Related Work}
As GenAI rapidly enters everyday software development, understanding existing productivity measurement frameworks and prior GenAI studies has become a key focus in SE research.

\vspace{-3mm}
\subsection{Frameworks for Measuring Productivity}
\boldification{Developer productivity is a complex multidimensional concept}
Developer productivity is a complex, multifaceted concept that has been studied by both academics and practitioners~\cite{peng2023impact}. Traditional metrics such as lines of code (LoC) \cite{mockus2002two}, commit counts \cite{vasilescu2015quality}, and task completion rates \cite{zhou2010developer} capture only narrow aspects of work and can be misleading or easily gamed. For example, the number of commits is often inflated by shallow or trivial changes \cite{oliveira2020code}, and LOC measures may reward verbosity rather than efficiency. Speed and volume of activity alone do not present a complete picture of productivity. Other forms of work—such as design, problem solving, and knowledge sharing—are equally important \cite{forsgren2021space}, as are software quality, impact, and delivery speed \cite{storey2022developers}.

%Emerging research, therefore, is investigating multi-dimensional productivity measurement frameworks.

\boldification{Multifaceted productivity measurement frameworks are...}
One such framework is the Developer Experience (DevEx) framework \cite{noda2023devex}, which highlights feedback loops, cognitive load, and flow state as key drivers of developer effectiveness. At the system level, the DORA (DevOps Research and Assessment) metrics \cite{wilkes2023framework} operationalize software delivery performance through deployment frequency, change lead time, mean time to recovery, and change failure rate. While these frameworks broaden the perspective beyond code quantity, they primarily address specific contexts of individual experience (DevEx) and delivery throughput (DORA).

\boldification{SPACE}
\citet{forsgren2021space} proposed the SPACE framework, a model that conceptualizes productivity as a combination of interpersonal and technical dimensions. The framework emphasizes that productivity arises from the interplay among human, technical, and organizational factors. For instance, high \textit{activity} (e.g., frequent commits) does not necessarily indicate improved \textit{performance} unless accompanied by quality, sustained satisfaction, collaboration, and flow. In this study, we adopt the SPACE framework as an analytical lens to examine how GenAI adoption influences developer productivity across its five dimensions: \textbf{S}atisfaction and well-being, \textbf{P}erformance, \textbf{A}ctivity, \textbf{C}ommunication and collaboration, and \textbf{E}fficiency and flow. 

\textit{Satisfaction and well-being} captures how fulfilled, motivated, and supported developers feel in their work. \textit{Performance} represents the quality and impact of outcomes (e.g., software reliability, feature completeness, or user satisfaction). \textit{Activity} refers to the volume of work performed (e.g., commits and code reviews). \textit{Communication and collaboration} refers to how developers communicate and work together. \textit{Efficiency and flow} describes how smoothly and continuously work proceeds (e.g., ability to maintain focus and minimize interruptions). 

Critically, the SPACE framework views productivity as a system of interdependent dimensions rather than isolated metrics. High activity without corresponding performance gains may indicate redistributed rather than reduced effort. Similarly, efficiency gains that increase cognitive load trade one constraint for another. This systems view positions SPACE as both a measurement and planning framework for understanding how interventions affect the full productivity landscape.

\vspace{-3mm}
\subsection{Empirical Evidence on Developer Productivity with GenAI}
As GenAI tools are increasingly integrated into software development, recent studies have highlighted productivity gains through faster task completion. For example, \citet{dohmke2023sea} reported that developers using GitHub Copilot completed programming tasks up to 55.8\% faster. In a randomized controlled trial with 96 full-time Google engineers, \citet{paradis2025much} found that AI assistance reduced time spent on a complex, enterprise-grade coding task by approximately 21\%. Additionally, \citet{rodriguez2023codequality} found that GenAI accelerated task completion and streamlined code review processes by suggesting actionable improvements and reducing reviewer effort.

However, recent literature suggests that GenAI’s productivity benefits may be overstated. AI-generated code can be inconsistent or incorrect, often requiring developers to re-prompt, validate, and debug the output, which then interrupts their workflow and reduces overall efficiency \cite{miller2025maybe, becker2025measuring}. \citet{kuhail2024will} found that over-reliance on AI assistance may erode developers’ coding proficiency and critical thinking, while reducing peer interaction, such as asking fewer questions, sharing fewer solutions, and engaging less in coordination \cite{tobisch2025knowledge, khojah2024beyond, qiu2025today}. Moreover, the 2025 Stack Overflow Survey further reported that only 29\% of developers trusted AI accuracy and 66\% spent more time debugging than expected \cite{Shimel2025AIAdoption}.

These studies offer fragmented views of GenAI’s impact on software development, focusing on programming tasks and short-term efficiency gains while overlooking broader productivity gains. This paper presents a comprehensive analysis of developer productivity in GenAI-mediated development, grounded in the multi-dimensional \emph{SPACE} framework.
\vspace{-3mm}

%% file: Tables/survey_mapping.tex
\begin{table}[]
\caption{Survey Items mapping}
\vspace{-6pt}
\resizebox{0.45\textwidth}{!}{
\begin{tabular}{ll}
\hline
\textbf{} & \textbf{Survey Items} \\
\hline
 \multicolumn{2}{c}{\textbf{Satisfaction and well-being}} \\
\hline
S1 &  I feel my current workload is manageable                                     \\
\rowcolor[HTML]{EFEFEF} 
S2 &  I feel mentally and physically exhausted from work $(r)^*$                           \\
S3 &  I am becoming less interested in work $(r)^*$                                        \\
\rowcolor[HTML]{EFEFEF} 
S4 &  I feel secure in my job and confident about my future in this company        \\ \hline

\multicolumn{2}{c}{\textbf{Performance}} \\
\hline
P1 &  The number of lines of code I change per day                                 \\
\rowcolor[HTML]{EFEFEF} 
P2 &  The proportion of my test cases that pass                                    \\
P3 &  The number of new API methods I learn each day                            \\ \hline

\multicolumn{2}{c}{\textbf{Activity}} \\
\hline
A1 &  The number of commits I make                                                    \\
\rowcolor[HTML]{EFEFEF} 
A2 &  The amount of time I devote to writing code                                  \\
A3 &  The number of test cases I write                                             \\
\rowcolor[HTML]{EFEFEF} 
A4 &  The amount of time I spend browsing the web for work-related information $(r)^*$  
                    \\
A5 &  The number of work items (tasks, bugs) I close      
                   \\
\rowcolor[HTML]{EFEFEF} 
A6 &  The number of code reviews I contribute to                                   \\
A7 &  The amount of time I devote to reviewing code $(r)^*$         
\\ \hline 

\multicolumn{2}{c}{\textbf{Communication and collaboration}} \\
\hline
C1 &  The number of meetings I attend $(r)^*$                                           \\
\rowcolor[HTML]{EFEFEF} 
C2 &  The amount of time I spend in meetings $(r)^*$                                         \\
C3 &  The number of work-related emails I write                                    \\
\rowcolor[HTML]{EFEFEF} 
C4 &  The amount of time I spend responding to email$(r)^*$                                    \\ \hline

\multicolumn{2}{c}{\textbf{Efficiency and flow}} \\
\hline
E1 &  The amount of time I spend on each work item $(r)^*$   \\
\rowcolor[HTML]{EFEFEF} 
E2 &  The amount of time I spend browsing the web for personal matters during work $(r)^*$
\\
\hline
\end{tabular}
}

% \raggedright
{\footnotesize *(r) indicates the items reverse-coded for visualization.}
\label{tab:survey_items}
\end{table}

%% file: sections/3_Methodology.tex
\section{Research Method}
To capture developers’ multi-dimensional productivity perceptions with genAI, we used a survey-based approach grounded in the SPACE framework \cite{forsgren2021space}.
% , adapting validated measures to the GenAI context.
\vspace{-3mm}
\subsection{Survey Design}
We designed our survey instrument to comprise the five dimensions of the  SPACE framework for productivity measurement \cite{forsgren2021space}. Since at the time of the survey design there was no validated questionnaire for SPACE, we adapted the questionnaires from \citet{meyer2014software, trinkenreich2024predicting, Casic_Panselina_2025}, which measured developer satisfaction, activity, performance, collaboration, and efficiency. The adaptations were made to reflect the GenAI context. See the complete questionnaire in the supplementary material~\cite{supply}. Respondents were offered the option to enter a \$50 raffle for every 50 responses. The protocol was approved by the university's IRB.

\noindent \textbf{Survey Questions}: We first asked questions about participants' demographics, including gender, professional background (role, years of experience, seniority level), and organization size. We also collected information about genAI-usage experience (frequency of use, adapted from \cite{usage2025}) to assess its perceived impact on productivity. The rest of the survey focused on each of the dimensions in the SPACE framework:

%Satisfaction
\noindent\textit{\textbf{Satisfaction} and well-being} measures developers' perception of fulfillment, motivation, and support about their work and team \cite{forsgren2021space}. Four statements assessed (1) workload manageability, (2) physical and mental exhaustion, (3) engagement, and (4) job security~\cite{trinkenreich2024predicting, Casic_Panselina_2025}.

%is how fulfilled developers feel in their work and team, as well as their overall health and happiness \cite{forsgren2021space}. This dimension was measured using four statements capturing developers' (1) workload manageability, (2) physical and mental exhaustion, (3) engagement at work, and (4) feeling of job security. These items were adapted from prior research \cite{trinkenreich2024predicting, Casic_Panselina_2025}.

%Performance
\noindent\textit{\textbf{Performance}} captures perceived outcomes of the development process, such as velocity~\cite{forsgren2021space}. We used three items adapted from \citet{meyer2014software} to assess the impact of AI on: (1) number of lines of code changed per day, (2) proportion of test cases passed, and (3) number of API methods learned.

% 4 - Activity
\noindent\textit{\textbf{Activity}} 
presents the volume of work performed \cite{forsgren2021space}, measured through seven statements adapted from \citet{meyer2014software}: (1) number of commits, (2) time spent writing code, (3) number of test cases, (4) time spent doing work-related browsing, (5) number of closed work items, (6) number of code reviews, and (7) time spent reviewing code.

%Comm
\noindent\textit{\textbf{Communication} and collaboration} describes how developers interact and coordinate~\cite{forsgren2021space}. We adapted four statements from \citet{meyer2014software}: (1) number of meetings attended, (2) time spent in meetings, (3) number of work-related emails written, and (4) time spent responding to emails.

\noindent\textit{\textbf{Efficiency} and flow} captures making progress with minimal interruptions \cite{forsgren2021space}. We assessed this by using two adapted statements from \citet{meyer2014software}: (1) time spent per work item and (2) time spent on personal browsing during work.

% final open-ended question
Finally, we added an open-ended question inviting participants to share reflections on how AI adoption has influenced their work. The survey took approximately 5–8 minutes to complete, and we held a raffle for a \$50 gift card, which participants could enter by sharing their email.

% sandbox and pilot survey
\vspace{3pt}\noindent\textbf{Sandbox and Pilot Survey.} We sandboxed the survey with five people with software engineering research experience, iteratively refining it until no further concerns arose. We then piloted with two professionals: one from a multinational organization and one from a U.S. national lab; both found the survey clear, easy to follow, and relevant to assess productivity.

% participant recruitment
\vspace{3pt}\noindent\textbf{Participant Recruitment.} Following ethical data collection guidelines, we recruited participants from diverse technical backgrounds, expertise, organization sizes, and domains. We recruited participants from 56 OSS communities, including organizational repositories (e.g., IBM, Oracle, Google), widely-used infrastructure, AI projects, and data science communities (e.g., PyTorch). We invited participants via email with study details and consent information. Responses were anonymized per GDPR and IRB approval. This two-week survey followed established SE research practices \cite{feng2025domains, feng2025multifaceted, feng2022case}.

% \input{Tables/Participant_distribution}

% participant distribution
%\textbf{Participant Distribution.} 
We received 688 responses. After removing invalid entries, we kept 415. Most participants worked in full-stack (36.9\%), backend (16.9\%), or data/ML roles (15.4\%). A majority identified as men (90.6\%), over half worked at large or extra-large organizations (57.8\%), and most (82.2\%) had more than five years of experience. More details in the supplementary material~\cite{supply}.

\vspace{-3mm}
\subsection{Data Analysis}
\textbf{RQ1.Productivity Impacts of GenAI Adoption.}
To answer RQ1, we analyzed how productivity perceptions vary by GenAI usage frequency. We categorized participants based on their response to \textit{``how often do you use AI tools for your software development work?''} into non-frequent (Never, Rarely, Sometimes) and frequent users (Often, Always).

We analyzed the Likert-scale response distributions and trends using stacked bar charts and violin plots. % to compare the two groups. 
The analysis focused on the five dimensions of the SPACE framework: Satisfaction and well-being (S1–S4), Performance (P1–P3), Activity (A1–A7), Communication and collaboration (C1–C4), and Efficiency and flow (E1–E2). To improve interpretability, we reverse-coded selected items (e.g., S2 and S3). Table~\ref{tab:survey_items} presents the mapping of the survey items. The complete set of charts is available in the supplementary material~\cite{supply}.

\noindent\textbf{RQ2. Perceived Productivity Gaps, Challenges, and Strategies.}
To understand developers' experiences with GenAI adoption and to identify perceived gaps, challenges, and potential strategies, we addressed RQ2 through a qualitative analysis of responses (N = 206) to the open-ended question: \textit{``If you have any final thoughts or experiences on how adopting AI tools has impacted your work, please share them below.''}

We used the five dimensions of SPACE as our codeset. Two authors independently coded 20\% of the responses, and compared their outputs using the Jaccard index \cite{landis1977measurement}, achieving 90\% agreement on inter-rater reliability. Given this level of consensus \cite{landis1977measurement}, one of the researchers coded the remaining responses. Based on this analysis, we developed a empirical mapping that maps seven identified gaps/challenges and eight potential strategies onto the SPACE dimensions.

%% file: sections/4_Results.tex
\section{Results}
Section \ref{sec:rq1_results} presents the results for RQ1, examining how GenAI adoption impacts different dimensions of developer productivity through the SPACE framework, highlighting where productivity improves and where it remains unchanged. Section \ref{sec:rq2_results} reports the results for RQ2, identifying the remaining gaps and challenges in using GenAI, and the potential strategies to address those. 

% Together, these results provide a nuanced view of both the productivity gains enabled by GenAI and the limitations that continue to shape developers' experiences. 

Our results show that GenAI adoption creates a constraint redistribution problem. Section \ref{sec:rq1_results} documents how GenAI-facilitated improvements in Activity and Efficiency dimensions create demands in Satisfaction, Performance, and Communication dimensions. Section \ref{sec:rq2_results} identifies specific constrain patterns and strategies that teams planning GenAI adoption or GenAI tool developers can use to anticipate these tradeoffs.

\vspace{-3mm}
\subsection{Productivity Across SPACE Dimensions (RQ1)}
\label{sec:rq1_results}
To answer RQ1, we begin by examining overall trends across the SPACE framework’s dimensions, and then drill down into item-level variations within each dimension.

\begin{figure*}[t]
    \centering
    \captionsetup{skip=10pt} % reduces space between image and caption
    \setlength{\belowcaptionskip}{-10pt}
    \includegraphics[width=\textwidth]{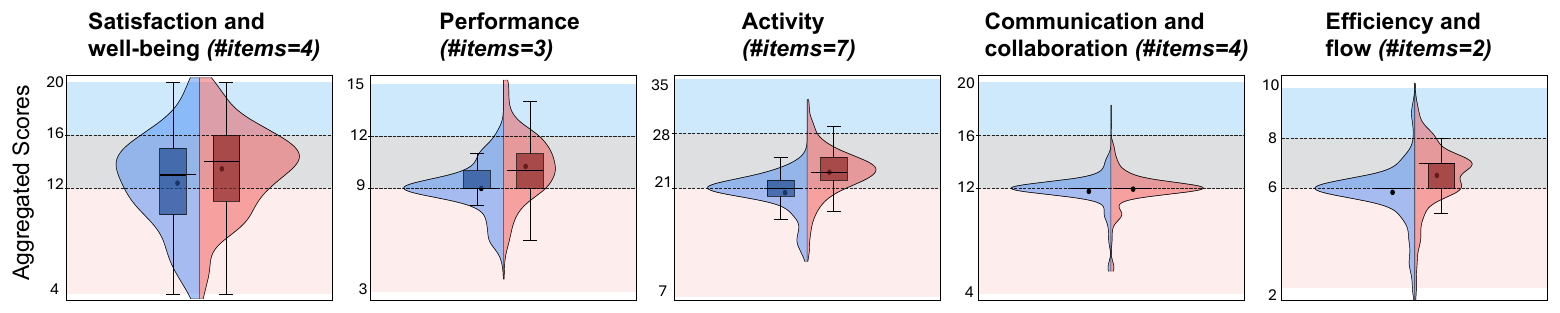}
   % \vspace{-10pt}
   
    \caption{Violin plots of aggregated SPACE scores. Left = non-frequent AI users (blue), right = frequent AI users (red). Boxplots show median (line) and mean (dot). Gray band marks the neutral (``no-change''); blue marks positive and red negative perceptions; example tick labels shown for \textit{Satisfaction \& Well-being} (4 items: 12 = all Neutral, 16 = all Agree).}
    \label{fig:violin-space}
\end{figure*}

Figure \ref{fig:violin-space} displays the median values for non-frequent and frequent AI user groups across the SPACE dimensions. Frequent AI users have slightly higher median scores than non-frequent users in all dimensions except \textit{Communication and collaboration}, where medians are the same. Nevertheless, all values remain \textit{within the neutral (or ``no-change'') range}, as indicated by the gray-shaded area. These results suggest that GenAI adoption has not yet produced substantial positive or negative shifts in productivity.

These overall patterns indicate that, at an aggregate level, GenAI adoption has not yet led to pronounced positive or negative shifts in developers’ perceived productivity. Instead, the effects appear incremental and uneven, with small median differences that may mask meaningful variation at the item level. Accordingly, we next examine each SPACE dimension in detail to unpack where GenAI adoption is associated with concrete improvements, where effects are limited, and where challenges persist across specific productivity aspects. Refer to the supplementary item-level stacked bar charts (Figs. 1–5) for each dimension of SPACE.

\emph{\textbf{Satisfaction and well-being.}} Among frequent AI users, positive ratings increased for workload manageability ($S1$) and perceptions of job security ($S4$): 68.6\% and 55.4\% \emph{Agreed/Strongly Agreed}, respectively. A similar pattern was observed among non-frequent AI users, 55.3\% ($S1$) and 52.9\% ($S4$) reported positive perceptions on these items.

Developers described these gains as GenAI easing repetitive tasks in their work, \textit{``I no longer need to waste my time with repetitive tasks. There is always a way to automate with AI. [P369]''}, \textit{``AI tools have had an overall positive impact on my work satisfaction; fewer mundane tasks, more focus time. [P3]'' }

However, these improvements are partial; more than half of respondents still reported feeling exhausted ($S2$: 65.2\% vs. 62.8\%). Participants noted that efficiency gains can accompany expectations to deliver more and faster, particularly in contexts that overestimate AI’s capabilities. As one participant observed, \textit{``There's also a growing expectation to deliver more simply because these tools are available, and sometimes non-technical stakeholders view AI as a ‘magic ball’ that can solve everything.''} Around half of the participants became less interested in work ($S3$: 46\% for frequent and 59.6\% for non-frequent users). \textit{``I move fast with AI and move mountains of work, but I am losing my passion for the craft [P201]''}.

\begin{spacebubble}{Satisfaction and well-being (S)}
  \begin{spacepositive}
    \bitem More developers report manageable workloads.
    \bitem Increased confidence in job security.
    \bitem Reduced disinterest in daily work.
  \end{spacepositive}
  \begin{spacenegative}
    \bitem High levels of exhaustion persist despite AI adoption.
  \end{spacenegative}
\end{spacebubble}

\emph{\textbf{Performance.}} Non-frequent AI users predominantly reported \emph{no change} in work quality or outcomes across performance-related items ($P1$: 66.7\%, $P2$: 83.1\%, $P3$: 66.7\%). In contrast, among frequent users, coding throughput ($P1$) increased: 72.7\% (47.2\% \emph{More} + 25.5\% \emph{Much more}). However, these gains did not extend uniformly to other performance items. For test case pass rates ($P2$) and learning velocity (API methods learned per day, $P3$), most of the frequent AI users---67.4\% and 58.6\% respectively---reported \emph{no change} or a decline. 

Qualitative responses help explain this divergence. While AI-driven acceleration allows developers to deliver more lines of code and generate broader test coverage, increased volume often shifts effort to review and validation, limiting net performance gains. 
\textit{``When I use code that's been written with AI, I feel slightly less control over my outcomes, even with through unit tests [P138]''}. Besides, the absence of oversight may constrain genuine performance improvement, highlighting that \textit{speed does not always equal progress}. \textit{``By my estimate, LLM-based tools have so far been more about speed than quality improvement [P5].''}

\begin{spacebubble}{Performance (P)}
  \begin{spacepositive}
    \bitem Higher coding throughput with frequent GenAI use.
    \bitem Greater volume of code changes delivered per day.
  \end{spacepositive}
  \begin{spacenegative}
    \bitem Test success rates show little to no improvement.
    \bitem Learning velocity remains largely unchanged.
  \end{spacenegative}
\end{spacebubble}

\emph{\textbf{Activity.}} In most activity-related items, frequent AI users reported higher levels of activity than non-frequent users, except code review ($A6$ and $A7$). Frequent AI users were less likely to report spending more time writing code ($A2$) or searching for work-related information ($A4$), with 75.4\% and 61.3\% of them, respectively, reporting no increase. Moreover, a larger share of frequent users reported producing more commits ($A1$): 48.3\% vs 7.9\%; more test cases ($A3$): 56.5\% vs 24.8\%; and completing more work items ($A5$): 55.9\% vs 9.1\%, compared to non-frequent users. \textit{``Agentic systems streamline small coding tasks and free up time for design thinking. [P1]'' } and \textit{``I spend less time on generating the code, and more time evaluating its behavior and design [P104].''}

However, these activity gains are accompanied by increased review effort. The majority of frequent users (84.3\%) reported that GenAI did not reduce the time spent on code reviews ($A7$), and frequent users were more likely than non-frequent users to report conducting more code reviews ($A6$: 25.1\% vs. 9.8\%). As one participant noted, \textit{``AI-generated code unfairly puts more onus on code reviewers to understand how the code works and find bugs or security issues.[P204]''} Developers emphasized that the acceleration often shifts effort toward validation and maintenance. \textit{``Reviewing LLM-generated content such as code and docs wastes time — coworkers are (accidentally but carelessly) sabotaging our work by `creating work'.The LLMs save these coworkers time because they are faster at producing content, but other coworkers have to spend disproportionately more time to review and correct all that content [P127].''} % \textit{``AI can give you a productivity boost by bringing you 70\% there in a few minutes. However, the last 30\% will likely take you close to what you would’ve done before because you also have to thoroughly review AI code. [P20]''}} 
Despite this shift, developers viewed the trade-off favorably sometimes, noting that \textit{``adopting AI tools has greatly improved my workflow, especially by enabling the faster creation of tests and use cases. While I do spend more time on projects now, it's still far less than it would take to create everything manually, and the overall quality and coverage have improved significantly.[P338]''}

\begin{spacebubble}{Activity (A)}
  \begin{spacepositive}
    \bitem Increased output of commits, test cases, and completed work items.
    \bitem Reduced time on direct code writing.
    \bitem Reduced time searching for work-related information.
  \end{spacepositive}
  \begin{spacenegative}
    \bitem Increased involvement in code review activities.
    \bitem No reduction in time spent reviewing code.
  \end{spacenegative}
\end{spacebubble}

\emph{\textbf{Communication and collaboration.}} 
% More than 75\% of developers fall on the non-positive side of the scale (i.e., reporting \emph{No Change} across all communication and collaboration items ($C1$–$C4$), regardless of AI usage frequency.
Across all communication and collaboration items ($C1$–$C4$), most of the developers reported no positive change, regardless of AI usage frequency. Specifically, more than three-quarters of all users reported no positive change (i.e., reporting \emph{No Change}) for meeting frequency ($C1$: 93.1\% vs 90.8\%), time spent in meetings ($C2$: 92.3\% vs 90.1\%), work-related emails written ($C3$: 93.6\% vs 91.3\%), and time spent responding to emails ($C4$:76.3\% vs 82.3\%). Among frequent AI users, \emph{No Change} responses dominate across all four items, exceeding 70\% in each case, indicating that GenAI adoption has not substantially altered day-to-day communication practices. 

However, participants noted GenAI reducing friction in personal communication tasks: \textit{``Superhuman’s AI model lets me search my inbox and answer questions related to email with very little time and effort. This saves at least an hour a week.[P287]''} 

\begin{spacebubble}{Communication and Collaboration (C)}
  \begin{spacenegative}
    \bitem Team communication and collaboration patterns remain largely unchanged with GenAI use.
    \bitem Meetings and email-related activities show little to no reduction.
  \end{spacenegative}
\end{spacebubble}

\emph{\textbf{Efficiency and flow.}} Frequent AI users were less likely than non-frequent users to report spending more time on each work item ($E1$: 35.8\% vs 82.2\%). Developers attributed this gain to AI's ability to accelerate a wide range of knowledge and content creation tasks: \textit{``By leveraging AI across various tasks such as writing, ideation, presentation creation, coding, information gathering, and research, I have been able to achieve significant results in a very short period of time. [P180]'' } and \textit{``AI has improved my efficiency by helping me not spend time checking syntax so I can focus on functionality [P321].''} 

Frequent AI users also reported spending less time than non-frequent AI users on personal web browsing during work ($E2$: 19\% vs. 7.1\%). However, the fact that 76.3\% of frequent AI users reported non-positive responses regarding reductions in interruptions indicates that GenAI use alone does not mitigate context switching or external disruptions. 

\begin{spacebubble}{Efficiency and Flow (E)}
  \begin{spacepositive}
    \bitem Reduced time on individual work items.
    \bitem Reduced time on non-work-related web browsing.
  \end{spacepositive}
  \begin{spacenegative}
    \bitem Improvements in sustained focus and flow are limited.
  \end{spacenegative}
\end{spacebubble}

% \vspace{-3mm}
\subsection{Productivity Challenges and Potential Strategies (RQ2)}
\label{sec:rq2_results}

\input{Tables/RQ2_Table}

To answer RQ2, we qualitatively analyzed open-ended responses using the SPACE dimensions to identify the gaps and challenges underlying the observations reported in the RQ1 results (highlighted in red in Section~\ref{sec:rq1_results}), as well as perceived strategies to mitigate them. Table~\ref{tab:AI_challenges_strategies} shows an empirical mapping of these challenges and the corresponding strategies.

\textbf{Satisfaction and well-being.}
Persistent high levels of exhaustion despite GenAI adoption emerged as the primary negative observation in the RQ1 results. One contributing factor is the \textcolor{spacered}{cognitive load} (\textit{Ch1}) associated with repeated prompting, verification, debugging, and error correction of AI outputs. \textit{``I'm able to close our features faster, but for the cost of significantly higher human brain compute load. [P207]''} This also includes  \textcolor{spacered}{ review burden} (\textit{Ch2}) of others' AI generated code. \textit{``...I was feeling resentment because I was sure I was spending more time reviewing and re-reviewing her doc than she had spent working on it.[P117]''} Moreover, the perception that GenAI accelerates output generation can raise organizational expectations, resulting in \textcolor{spacered}{ increased pressure and stress on developers} (\textit{Ch3}). \textit{``AI tools so far have been similar to other types of tools in practice... but with much more management hype / misplaced expectations and general feelings of dread [P160]''}.

\textcolor{spaceblue}{Surfacing uncertainty through confidence indicators and explanation norms} (\textit{St3}) can help developers calibrate trust in GenAI outputs better, and reduce review burden. For example, when submitting a pull request, developers could be required to indicate which parts were AI-generated, their confidence in the correctness of those changes, and any verification steps taken. As one participant noted, \textit{``Junior developers don't know how to use it and they blindly apply it without thinking it through before submitting. [P234]''}  Additionally, because organizational norms strongly shape how GenAI affects well-being, teams and management should foster an environment that treats  \textcolor{spaceblue}{GenAI as assistive rather than as a mechanism to push for increased output} (\textit{St2}).\textit{``Having training and clear policy for use has really helped our team to adopt the tools in a positive manner.[P268]''} Subsequently, \textcolor{spaceblue}{ organizational-level training} (\textit{St1}) and clear usage policies can support more positive adoption. \textit{``My company has made broad statements encouraging us to use AI to help us be more productive, but has never offered any training on how to do it.[P351]''}

\textbf{Performance.}
Due to the \textcolor{spacered}{verbosity of AI-generated outputs} (\textit{Ch4}), particularly in test generation, test success rates may show little to no improvement. \textit{``AI generates many test cases, including verbose ones, which makes code-reviewing more time-consuming [P23].''} Accordingly, \textcolor{spacered}{reliance on AI for tasks before acquiring sufficient foundational knowledge} (\textit{Ch5}) contributes to unchanged learning velocity. Developers were concerned that relying on GenAI too early can bypass key learning processes, especially for junior developers: \textit{``I fear that juniors will take much longer to get to an experience level that is comparable to what devs with 10+ years of experience. [P192]''}

One potential strategy to mitigate challenges related to test quality is to \textcolor{spaceblue}{ integrate GenAI tools with project-specific context} (\textit{St4}), including providing access to design decisions, architectural design, and broader project context, so that AI-generated tests better reflect system intent rather than generic assumptions based on small chunk of code. \textit{``We only use AI as if it were a ‘genius graduate’ junior coder that does not understand business needs and workflow. [P14]''}

Developers highlighted the importance of \textcolor{spaceblue}{ pairing GenAI use with structured fundamental learning resources} (\textit{St5}) rather than relying on AI as a substitute for fundamental skill development in situations where the task demands knowledge beyond their existing expertise. \textit{``What it concerns me is that when I ask it to write code in languages I have a lot of expertise in, it's all bad and I hate it, so that makes me a lot less confident in the code it generated for languages I know a lot less well.[P57]''} \textit{``I'd prefer my teammates embrace investing in themselves and learning fundamentals instead of relying on AI to produce sub-par code.[P415]''}

\textbf{Activity.}
With GenAI adoption, developers reported increased involvement in code reviews without a corresponding reduction in workload, reflecting a shift in effort rather than a net savings. In particular, \textcolor{spacered}{ AI shifts effort toward reviewing others' AI-generated outputs} (\textit{Ch2}), as higher output volumes increase supervision needs. \textit{``There's an increase in the amount of effort and time I'm reviewing work of other folks cranking out more stuff with the help of AI tools.[P359]''} This burden is further compounded by the \textcolor{spacered}{ verbosity of AI outputs} (\textit{Ch4}), which often results in low-quality content. \textit{``...Reviewing that low-quality content and giving feedback on it consumes a lot of time [P127].''} \textit{``Sometimes you spend more time reviewing than if you had written it yourself.[P207]''}

Using AI as a \textcolor{spaceblue}{first-pass reviewer rather than as a source of final, review-ready content} (\textit{St6}) may help reduce review burden.\textit{``Reviewing code written by AI is painful, when the author did not take care to clean it up. [P355]''}. Additionally, as discussed earlier, \textcolor{spaceblue}{ surfacing uncertainty via confidence indicators and explanation norms} (\textit{St3}) could prevent additional supervision work from being pushed downstream. \textit{``AI tools have shifted me into a role where I spend more time performing detailed review of code [P104]''}. Together, these observations highlight the importance of introducing \textcolor{spaceblue}{ quality gates for AI-heavy changes} (\textit{St7}) to mitigate review overhead. \textit{``LLM approach needs strict guidance and ideal examples to mimic. If your code is sloppy or undocumented, your results will probably be poor. It's just a fuzzy mirror.[P87]''}

\textbf{Communication and collaboration.} Our RQ1 findings indicate that Communication and collaboration showed little to no change with the adoption of GenAI. We do not frame outcome of this as a challenge; instead, it highlights an explanatory gap regarding where GenAI’s impact is limited. A potential explanation is that GenAI tools currently emphasize \textcolor{spacered}{support for individual work activities, offering less direct support for shared coordination and collective practices} (\textit{Ch6}).
Developers’ experiences suggest that while GenAI can assist with individual productivity, it does not substantially alter how teams align or coordinate work. This aligns with prior empirical findings showing that AI tools tend to optimize individual task execution but have limited impact on collaborative practices such as meetings and coordination routines \cite{dillon2025shifting}.

Moreover, \textcolor{spacered}{collaboration often depends on human alignment} (\textit{Ch7}) including shared understanding, and social negotiation, which are difficult to automate through GenAI alone \cite{woolley2025generative}. \textit{``Creativity gets a collaborator, not a replacement. [P269]''}

Therefore, developers and prior work pointed toward organizational strategies that clarify GenAI’s role in collaborative settings. First, setting team- and management-level norms that frame \textcolor{spaceblue}{ GenAI as assistive} (\textit{St2}) rather than as a replacement  may help prevent over-reliance on AI at the individual level while preserving human-centered collaboration.

In addition, \textcolor{spaceblue}{ defining guidelines for GenAI use in shared communication artifacts} (\textit{St8}) emerged as a relevant strategy, aimed at making individual and ad-hoc GenAI use in communication work more explicit and collectively aligned \cite{unc2025genai}. \textit{``Our organization does use it more regularly for non-code tasks such as meeting notes/summaries, incident summaries, documentation, etc. [P75]''}  While prior studies suggest that GenAI may reduce time spent in less relevant meetings and lower effort devoted to non-critical email communication \cite{microsoft2024worktrend, nysscpa2024survey}, our findings indicate that such collaboration-related practices remained largely unchanged. One possible explanation is that GenAI use in shared communication artifacts currently lacks shared norms or guidance.

% such as emails, meeting notes, summaries, and documentation can help teams benefit from AI support \cite{unc2025genai}. 

\textbf{Efficiency and flow.}
Developers faced challenges related to \textcolor{spacered}{ AI-induced cognitive workload from repeatedly verifying and correcting outputs} (\textit{Ch1}), which may undermine their ability to maintain concentration and result in limited improvements in sustained focus and flow. \textit{``I move fast with AI and move mountains of work, but I am losing my passion for the craft and the ability to quickly focus. [P201]''}

To mitigate this challenge, developers emphasized the importance of \textcolor{spaceblue}{ structured, organization-level training on how to use GenAI effectively} (\textit{St1}). Such training (e.g., prompt engineering) can help developers better scope problems, recognize when verification is necessary, and avoid inefficient prompting and rework. \textit{``If you have a very well-defined problem and some strong test cases, AI can really get there. [P20]''} \textit{``It massively improved certain refactors and updates, but required adapting approaches to tasks to ensure best usage of AI tools. [P57]''}

%% file: Tables/RQ2_Table.tex
\begin{table*}[ht]
\centering
\caption{The productivity challenges and potential strategies for the negative observations}
\vspace{-8pt}
\label{tab:AI_challenges_strategies}
\begin{adjustbox}{max width=0.9\textwidth}
\begin{tabular}{l P{4.8cm} P{7cm} P{7.2cm}}
\toprule\toprule
\textbf{SPACE} & \textbf{Observations} & \textbf{Challenges} & \textbf{Potential Strategy} \\
\midrule\midrule

\textbf{S} & High levels of exhaustion persist despite AI adoption. & 
\textbf{Ch1.} AI-induced cognitive workload from verifying outputs. \newline
\textbf{Ch2.} AI shifts effort to reviewing others' AI outputs rather than reducing total workload. \newline
\textbf{Ch3.} Organizational "AI = faster output" expectations increase pressure and stress.
& 
\textbf{St1.} Structured organizational training on using GenAI. \newline 
\textbf{St2.} Set team/management norms framing GenAI as assistive rather than a driver of higher output expectations. \newline 
\textbf{St3.} Surface uncertainty via confidence indicators and explanation norms.
\\
\midrule

 \textbf{P} & Test success rates show little to no improvement. \newline
 Learning velocity remains largely unchanged.
 & 
\textbf{Ch4.} Verbosity of AI outputs. \newline
\textbf{Ch5}. Reliance on AI for tasks before acquiring the necessary foundational knowledge
& 
\textbf{St4.} Integrate GenAI tools with project-specific context. \newline
\textbf{St5.} Pair GenAI use with structured fundamental learning resources. \\
\midrule

 \textbf{A} & Increased involvement in code review activities. \newline
 No reduction in time spent reviewing code. 
 & 
\textbf{Ch2.} AI shifts effort to reviewing others' AI outputs rather than reducing total workload. \newline
\textbf{Ch4.} Verbosity of AI outputs.
& 
\textbf{St6.} Use GenAI for first-pass review, not final approval. \newline 
\textbf{St3.} Surface uncertainty via confidence
indicators and explanation norms.\newline 
\textbf{St7.} Add quality gates for AI-heavy changes. 
\\
\midrule

 \textbf{C} & Team communication and collaboration patterns remain largely unchanged with GenAI use. \newline
 Meetings and email-related activities show little to no reduction.
 & 
\textbf{Ch6.} GenAI primarily supports individual tasks rather than shared coordination work. \newline
\textbf{Ch7.} Collaboration relies on human alignment rather than automatable tasks.
&
\textbf{St2.} Set team/management norms that treat GenAI as assistive. \newline
\textbf{St8.} Define guidelines for GenAI use in shared communication artifacts.
\\
\midrule

\textbf{E} & Improvements in sustained focus and flow are limited. & 
\textbf{Ch1.} AI-induced cognitive workload from verifying outputs. & 
\textbf{St1.} Structured organization level training on using GenAI.  \\
\bottomrule\bottomrule
\end{tabular}
\end{adjustbox}
\end{table*}

%% file: sections/5_Threats_to_validity.tex
% \vspace{-3mm}
\section{Limitations}
\textbf{Limitations of Perceptual Productivity Assessment.}
This study captures developers’ perceptions of GenAI’s impact on their productivity using self-reported survey data, which may not reflect objective productivity outcomes. However, we grounded our survey design in well-established productivity frameworks and prior validated measures used in related work \cite{meyer2014software, trinkenreich2024predicting, Casic_Panselina_2025}. Future studies are needed to complement perceptual findings with additional empirical analyses (e.g., repository mining or longitudinal behavioral measures).

\noindent \textbf{Limitations of Adoption Frequency as a Proxy.}
To address RQ1, we compared developers with different frequencies of GenAI use to investigate how adoption relates to productivity perception. While usage frequency provides a practical indicator of GenAI engagement, it may not fully capture other important dimensions of adoption, such as the types of tasks supported, the depth of integration into workflows, or reliance on specific GenAI features. To provide a more comprehensive perspective, we designed our analysis across multiple SPACE dimensions, and we complemented closed-ended items with an open-ended question to contextualize participants’ experiences. Future work should consider richer operationalizations of GenAI adoption (e.g., task-level or longitudinal behavioral measures) to further characterize its productivity impact.

\noindent \textbf{Limitations of Sample Representativeness.}
We acknowledge that no single sample can fully represent the global software workforce. However, our dataset includes 415 software practitioners from 56 organizations, which is comparable in scale and diversity to prior empirical studies of software engineers \cite{feng2025domains, trinkenreich2023belong}. Our participants span a broad range of company sizes, roles, and experience levels, offering diverse perspectives and providing a reasonable basis for understanding developers’ productivity perception in the context of GenAI adoption.

%% file: sections/6_discussion.tex
% \vspace{-3mm}
\section{Concluding Remarks}
GenAI tools are reshaping how developers engage with their work. Prior studies have demonstrated that these tools can accelerate task completion \cite{peng2023impact}. Yet our findings present a more complicated picture, that GenAI adoption has not yet produced the broad productivity gains that many organizations expect, but rather has led to ``spurious productivity,'' a surface-level acceleration that obscures stagnant or redistributed effort.

\vspace{-3mm}
\subsection{The Redistribution of Effort}
The concept of spurious productivity becomes evident when examining where developer effort shifts. Rather than eliminating work, developers described how AI tools redistributed it from code generation to downstream activities:

\noindent \textit{\textbf{Debugging of AI outputs.}} Developers described trading manual coding time for cycles of prompting, verifying, and fixing AI-generated code. \textit{``...any speed gained by code being produced faster is quickly lost when having to heavily scrutinize the generated code, and I have no interest in outsourcing the code I write to a AI tool only to become a professional code reviewer.[P161]''}

\noindent \textit{\textbf{Increased code review burden.}} Our data revealed that the increased volume of AI-assisted output created a corresponding increase in review workload. Frequent AI users were more likely than non-frequent users to report conducting a greater number of code reviews, and most (84.3\%) reported that GenAI did not reduce the time these reviews take. Participants explained that AI-generated code shifts cognitive burden to reviewers. \textit{``...I spend more time reviewing code anyway because a couple teammates have adopted AI tools. I have to review their code much more closely now because the output of these tools is frankly not very good, and the people using them are not paying close enough attention to the output.[P161]''} 

\noindent \textit{\textbf{Cognitive load from constant output verification.}} Beyond measurable review time, developers also described a less visible cost--- the mental effort required to continuously evaluate AI suggestions. This cognitive load may explain why exhaustion levels remained high despite reported efficiency gains in task completion. \textit{``In general, I welcome AI adoption, but the state of AI tools still produces very hit and miss outcomes and induces a ton of frustration.[P66]''}

\vspace{-3mm}
\subsection{Implications for Measuring Productivity}
% → Epistemic / evaluative: How should productivity be measured and interpreted?
% This is where Brooks fits naturally, and where you warn against activity-only metrics.

Our findings have important implications for evaluating GenAI adoption. While frequent AI users showed perceived increases in activity metrics such as more commits, test cases, and work items closed, performance metrics told a different story, with little improvement in test pass rates or learning velocity. Organizations that focus solely on activity-level measures may conclude that GenAI is effective, whereas those that assess performance outcomes may reach very different conclusions.

The disconnect recalls Fredrick Brooks' observation in the engineering management classic, \textit{The Mythical Man-Month} \cite{brooks1995mythical}, that there can be no single metric for programmer productivity, and that attempts to find one typically measure volume rather than performance. Organizations should use a holistic framework to evaluate productivity gains from AI. 

% This study shows how SPACE distinguishes Activity (volume of work performed) from Performance (quality and impact of outcomes) while accounting for redistributed effort.

\vspace{-3mm}
\subsection{SPACE as a Tool To Guide Practitioners}
% → Prescriptive / actionable: How should practitioners and tool builders act differently?
% This is where redistribution, constraints, and planning live.

The key insight of this study is that effort saved in one SPACE dimension often resurfaces in another. When AI generates code faster (Activity), verification demands increase (Efficiency and flow). When AI produces more output (Performance), the greater review burden increases cognitive load (Satisfaction and well-being). Practitioners and tool designers who focus on accelerating a single dimension without addressing the others will find that total effort remains roughly constant; it simply moves to wherever constraints are weakest.

Our work suggests that SPACE could provide a planning framework for this constraint problem. Consider a team adopting AI-assisted code generation. Rather than deploying the tool and measuring commits, the team could work backward through the SPACE dimensions to identify where effort will likely shift, then address those areas proactively. For example, before using AI to generate features, teams could use AI to create comprehensive test suites and design documentation to thoroughly constrain and validate the output. This investment in Performance artifacts might reduce the cognitive load required to verify AI-generated code later.

In \textbf{conclusion}, developers report only modest productivity gains from GenAI across \emph{SPACE}, with most dimensions showing little change and some improvements reflecting trade-offs or \emph{spurious} productivity. We also provide an empirical mapping of developers’ perceived challenges and potential strategies, offering actionable guidance for AI-infused workflows. Future work will complement these insights with repository mining and objective productivity signals to deepen the understanding of sustainable productivity under GenAI adoption.

 % AI-generated outputs. As a result, \emph{GenAI changes where developers spend their effort rather than increasing what they achieve.} GenAI tools are reshaping how developers engage with their work and accelerating task execution \cite{peng2023impact}. 